\documentclass[graybox]{svmult}

\usepackage{mathptmx}       
\usepackage{helvet}         
\usepackage{courier}        
\usepackage{type1cm}        
\usepackage{makeidx}         
\usepackage{graphicx}        
\usepackage{multicol}        
\usepackage[bottom]{footmisc}

 \usepackage[numbers]{natbib}
\usepackage{hyperref}

\makeindex             

\begin{document}

\title*{Archetypal cases for questionnaires with nominal multiple choice questions}
\author{Aleix Alcacer and Irene Epifanio}
\institute{Aleix Alcacer \at Universitat Jaume I, Castelló de la Plana, Spain \email{aalcacer@uji.es}
\and Irene Epifanio \at Universitat Jaume I and valgrAI, Spain \email{epifanio@uji.es}}
%
%
\maketitle

\abstract{Archetypal analysis serves as an exploratory tool that interprets a collection of observations as convex combinations of pure (extreme) patterns. When these patterns correspond to actual observations within the sample, they are termed archetypoids. For the first time, we propose applying archetypoid analysis to nominal observations, specifically for identifying archetypal cases from questionnaires featuring nominal multiple-choice questions with a single possible answer. This approach can enhance our understanding of a nominal data set, similar to its application in multivariate contexts. We compare this methodology with the use of archetype analysis and probabilistic archetypal analysis and demonstrate the benefits of this methodology using a real-world example: the German credit dataset.}

\keywords{Archetype analysis, Archetypoid analysis, Nominal observations, Dummy variables}

\section{Introduction}
\label{sec:1}
Nominal variables, which categorize characteristics into two or more distinct groups, are commonly found in research.  Nominal data are observed rather than measured, are unordered and non-equidistant, and do not possess a meaningful zero point. This type of data appears, for example, in surveys with multiple choice questions. In our case, participants can choose only one response from a predetermined set of choices.

For the first time, we introduce the exploratory tool Archetypoid Analysis (ADA) for this type of data to help understand, describe, visualize, and extract easily interpretable information, even for non-experts. ADA is an unsupervised statistical learning technique designed to approximate sample data through a convex combination (or mixture) of $k$ pure patterns known as archetypoids, which represent extreme observations within the sample. Their inclusion in the sample enhances interpretability, while their status as extreme cases aids in data comprehension. Research indicates that people grasp data more effectively when it is presented through its extreme components or when contrasting features of one observation against another \cite{Thurau12}. ADA was proposed by \cite{vinue2015archetypoids} as a derivative methodology of Archetype Analysis (AA), which was developed by \cite{cutler1994archetypal}. Like ADA, AA aims to approximate data using mixtures of archetypes for real continuous multivariate data; however, archetypes are not actual cases but rather combinations of data points. Furthermore, \cite{seth2016probabilistic} introduced a probabilistic framework for AA (PAA) to accommodate nominal observations \cite{EugsterPAMI} by operating within the parameter space. While AA and ADA were initially designed for real-valued observations, this work aims to extend archetypal tools to nominal data. For the particular case of dichotomous or binary variables, i.e. nominal variables which have only two categories, \cite{sort20} proposed using ADA instead of AA or PAA. For ordinal data, ADA was also used in a two-step methodology based on the ordered stereotype model \cite{FERNANDEZ2021281}.

In Section \ref{sec:2}, the methodology is explained, while data and results are
introduced in Section \ref{sec:3}. Finally, some conclusions are provided in Section \ref{sec:4}. The code and data is available at {\url{http://www3.uji.es/~epifanio/RESEARCH/nom.zip}}.

\section{Methodology}
\label{sec:2}
Nominal variables cannot be utilized in their categorical form within AA or ADA, since they are used with numeric variables. Dummy coding enables us to convert categories into a format that AA and ADA can manage, since character columns are converted into a serie of binary columns. As consequence, the first step consists of transforming the nominal matrix columns into dummy (binary) columns. In the second step, we apply ADA to this  $n \times m$ binary matrix, $\mathbf{X}$  with $n$ observations and $m$ columns.

In ADA, three matrices are defined: a) the $k$ archetypoids $\mathbf{z}_j$, represented as the rows of a $k \times m$ matrix $\mathbf{Z}$, which contains concrete rows of $\mathbf{X}$; b) an $n \times k$ matrix $\mathbf{\alpha} = (\alpha_{ij})$, which contains the mixture coefficients that approximate each observation $\mathbf{x}_i$ as a convex combination of the archetypoids ($\mathbf{\hat{x}}_i = \displaystyle \sum_{j=1}^k \alpha_{ij} \mathbf{z}_j$); and c) a $k \times n$ matrix $\mathbf{\beta} = (\beta_{jl})$, which holds the binary coefficients that define each archetypoid ($\mathbf{z}_j$ = $\sum_{l=1}^n \beta_{jl} \mathbf{x}_l$). To determine these matrices, we aim to minimize the following residual sum of squares (RSS), subject to specific constraints (where $\displaystyle \| \cdot\|$ denotes the Euclidean norm for vectors):

\begin{equation} \label{RSSar}
RSS = \displaystyle  \sum_{i=1}^n \| \mathbf{x}_i - \sum_{j=1}^k \alpha_{ij} \mathbf{z}_j\|^2 = \sum_{i=1}^n \| \mathbf{x}_i - \sum_{j=1}^k \alpha_{ij} \sum_{l=1}^n \beta_{jl} \mathbf{x}_l\|^2{,}
\end{equation}

with the following constraints:

1) $\displaystyle \sum_{j=1}^k \alpha_{ij} = 1$ with $\alpha_{ij} \geq 0$ {for} $i=1,\ldots,n$ {and}

2) $\displaystyle \sum_{l=1}^n \beta_{jl} = 1$ with $\beta_{jl} \in \{0,1\}$ and $j=1,\ldots,k$.

The estimation of the matrices in the ADA problem can be accomplished using the algorithm developed by \cite{vinue2015archetypoids}.

The concept behind archetypal analysis is that we can identify a set of archetypal patterns, with the understanding that data can be represented as a mixture of these patterns. In the context of nominal or binary data, it is essential that the archetypal patterns are also nominal or binary, reflecting the nature of the population from which the data originates. For instance, if pregnancy is one of the binary variables, it would be illogical to consider an archetypal observation as a woman who is pregnant 0.8 times. In other words, archetypal patterns must be nominal or binary to maintain clarity and interpretability, which are fundamental to archetypal techniques; they should represent observable phenomena rather than abstract concepts.

Additionally, to describe data as mixtures, we must assume that observations exist within a vector space, meaning they can be scaled (in this case, within the interval $\left[ 0, 1 \right]$) and combined. A suitable solution that encompasses these ideas is to apply ADA to $\mathbf{X}$, as the feasible archetypal patterns will belong to the observed sample. In fact, ADA was initially developed to address scenarios where genuine non-fictitious patterns were sought \cite{vinue2015archetypoids}. Conversely, the archetypes produced by applying AA or PAA do not necessarily need to be binary; they may not belong to the feasible set of solutions. Indeed, \cite{EugsterPAMI} binarized the archetypes obtained through AA or PAA in their experiments.

In fact, employing a continuous optimization approach for a problem where feasible solutions are not continuous can lead to significant failures \citep[Ch. 13]{Fletcher2000}. There is no assurance that this method will yield a satisfactory solution, even when considering all feasible binary solutions in proximity to the continuous solution. A simulation experiment involving binary observations was conducted by \cite{sort20}, who found that the archetypoids produced by ADA aligned more closely with the true archetypes compared to those generated by AA or PAA, in that order. Specifically, ADA resulted in the lowest mean misclassification error.

Consequently, we suggest utilizing ADA to manage nominal or binary observations. 
\section{Results}
\label{sec:3}
We examine the German credit data set utilized by \cite{EugsterPAMI}, which describes individuals and their credit applications. This data set is available at \cite{statlog_(german_credit_data)_144} and includes information on 1,000 people and five nominal variables (with the number of categories in parentheses): credit purpose (10), employment period (5), personal status and sex (4), job situation (4), and credit risk (2). The objective of this analysis is to identify archetypal profiles of credit applications. Firstly, we create dummy variables, and then ADA, AA and PAA are applied. Results for AA and PAA are binarized by replacing all values above 0.5
with 1 and others with 0. For comparative purposes, we use $k$ = 10, which was the value used by \cite{EugsterPAMI}.

\begin{figure}[b]
\includegraphics[scale=.5]{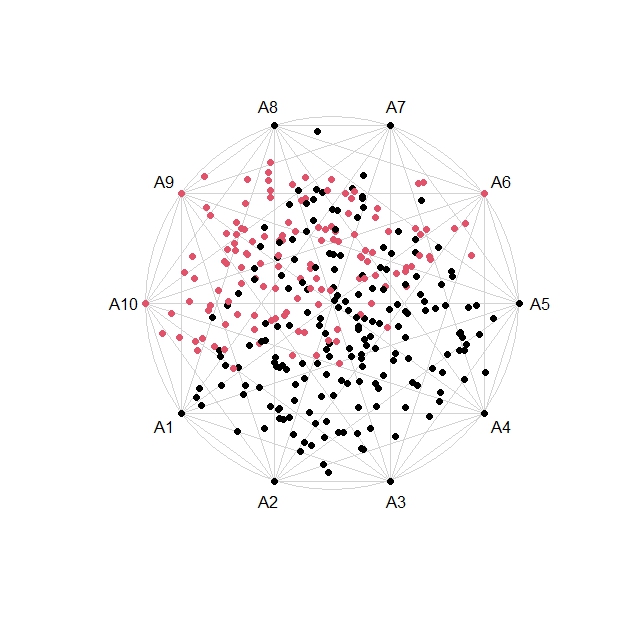}
%
%
\caption{Simplex visualization for the German credit dataset. Credit risk factor appeared as black (`good') and red (`bad').}
\label{fig:1}       
\end{figure}

Table \ref{tab:1} shows the archetypoids derived from ADA. Three of the profiles correspond with bad credit risk situation. One of the problems that can arise with AA and PAA is that no categories exceed the binarization threshold. In fact, this occurs with PAA for archetypal profile 6 and the variable credit purpose. Fig. \ref{fig:1} shows the alpha values projected through a simplex visualization as explained in \cite{EugsterPAMI}. The 'bad' observations are located around archetypoids 6, 9 and 10. 

To evaluate the complementarity of the archetypal profiles and therefore their extremity, we have computed the Hamming distance (the Manhattan distance between binary vectors) between the archetypal profiles. Table \ref{tab:2} displays a summary of the distances between the archetypal profiles of each method, along with the sum of those distances  in column `Total'. Archetypoids are thus able to identify more complementary profiles.

\begin{table}[]
\centering
\caption{Description of archetypoids obtained by ADA, together with the number of row selected as archetypoid (first column).}
\label{tab:1}       
\scriptsize
\begin{tabular}{lllp{0.2\columnwidth}p{0.3\columnwidth}c}
Case & Credit purpose & Employment period & Sex and personal status & Job situation & Credit risk \\ \hline
23  & car (new)           & ... \textless 1 year                  & male: single                       & unskilled - resident                                                  & Good \\
591 & radio/television    & 4  \textless{}= ... \textless 7 years & female  & unskilled - resident                                                  & Good \\
32  & furniture/equipment & 1  \textless{}= ... \textless 4 years & male: single                       & skilled employee/ official                                           & Good \\
15  & car (new)           & 1  \textless{}= ... \textless 4 years & female & skilled employee / official                                           & Good \\
129 & car (used)          & unemployed                            & male: single                       & self-employed/highly qualified staff& Good \\
125 & furniture/equipment & ...   \textless 1 year                & female & skilled employee / official                                           & Bad  \\
1   & radio/television    & .. \textgreater{}= 7 years            & male: single                       & skilled employee/ official                                           & Good \\
146 & business            & 4  \textless{}= ... \textless 7 years & male: single                       & skilled employee/ official                                           & Good \\
107 & car (new)           & .. \textgreater{}= 7 years            & male: single                       & self-employed/highly qualified staff & Bad  \\
436 & radio/television    & 1  \textless{}= ... \textless 4 years & male: married/widowed              & skilled employee/ official                                           & Bad 
\end{tabular}
\end{table}

\begin{table}[]
\centering
\caption{Number of entries with each Hamming distance between archetypal profiles for each method.}
\label{tab:2}       
\begin{tabular}{lllllllll}
Distance & 0  & 4  & 5  & 6  & 7 & 8  & 10 & Total \\ \hline
ADA    & 10 & 8  & 0  & 46 & 0 & 28 & 8  & 612   \\
AA     & 10 & 12 & 0  & 42 & 0 & 28 & 8  & 604   \\
PAA    & 10 & 12 & 12 & 26 & 6 & 28 & 6  & 590  
\end{tabular}
\end{table}
\section{Conclusions}
\label{sec:4}
We have proposed using ADA to identify archetypal patterns in nominal data, enhancing our understanding of a dataset. The results from a real-world application have demonstrated the advantages of ADA for questionnaires featuring nominal multiple-choice questions, serving as an alternative or complement to other established methodologies. While much of statistics relies on the principle that averaging across numerous elements in a dataset is beneficial, this paper takes a different approach. We focus on a small number of representative observations (archetypal observations) and explain the data composition through mixtures of these extreme cases. Our findings indicate that this method can be highly informative and serves as a valuable tool for making datasets more accessible and understandable, even for non-experts.

As future work, a not straightforward extension would involve addressing the scenario of mixed data, which includes both real-valued and categorical data, as well as missing values. Additionally, from a computational perspective, when dealing with very large datasets, the ADA algorithm proposed by \cite{vinue2015archetypoids} may be slow. In such cases, the method proposed by \cite{Vinue21} can be utilized for computing ADA with large datasets.

\begin{acknowledgement}
This work was supported by the Generalitat Valenciana CIPROM/2023/66, Spanish Ministry of Science and Innovation PID2022-141699NBI00
and PID2020-118763GA-I00. 
\end{acknowledgement}

\bibliographystyle{spbasic}
\bibliography{references}

@article{seth2016probabilistic,
  title={Probabilistic archetypal analysis},
  author={Seth, Sohan and Eugster, Manuel JA},
  journal={Machine Learning},
  volume={102},
  number={1},
  pages={85--113},
  year={2016},
  publisher={Springer}
}

@Book{Fletcher2000,
	title = {{Practical Methods of Optimization}},
	edition = {Second},
	year = {2000},
	publisher = {John Wiley \& Sons},
	author = {Fletcher, R.}
}

@article{FERNANDEZ2021281,
title = {Archetypal analysis for ordinal data},
journal = {Information Sciences},
volume = {579},
pages = {281-292},
year = {2021},
author = {Daniel Fernández and Irene Epifanio and Louise Fastier McMillan}}

@article{sort20,
    author     =     {Ismael Cabero and Irene Epifanio},
    title ={Finding archetypal patterns for binary questionnaires},
		journal ={SORT},
		year ={2020},
		volume ={44},
		number= {1},
		pages={39-66}
    }

@article{vinue2015archetypoids,
  title={Archetypoids: A new approach to define representative archetypal data},
  author={Vinu{\'e}, Guillermo and Epifanio, Irene and Alemany, Sandra},
  journal={Computational Statistics \& Data Analysis},
  volume={87},
  pages={102--115},
  year={2015},
  publisher={Elsevier}
}

@article{Vinue21,
  title={Robust archetypoids for anomaly detection in big functional data},
  author={Vinu\'e, Guillermo and Epifanio, Irene},
  journal={Advances in Data Analysis and Classification},
  volume={15},
  number={2},
  pages={437--462},
  year={2021},
  publisher={Springer}
}

@article{EugsterPAMI,
  author    = {Sohan Seth and
               Manuel J. A. Eugster},
  title     = {Archetypal Analysis for Nominal Observations},
  journal   = {{IEEE} Trans. Pattern Anal. Mach. Intell.},
  volume    = {38},
  number    = {5},
  pages     = {849--861},
  year      = {2016}
}

@article{cutler1994archetypal,
	title = {{Archetypal Analysis}},
	journal = {Technometrics},
	year = {1994},
	volume = {36},
	author = {Cutler, Adele and Breiman, Leo},
	pages = {338-347},
	number = {4}
}

@article{Thurau12,
year={2012},
journal={Data Mining and Knowledge Discovery},
volume={24},
number={2},
title={Descriptive matrix factorization for sustainability: Adopting the principle of opposites},
author={Thurau, Christian and Kersting, Kristian and Wahabzada, Mirwaes and Bauckhage, Christian},
pages={325-354}
}

\end{document}